\documentclass[aps,pra,twocolumn,superscriptaddress,showpacs]{revtex4-2}

\usepackage{graphicx}

\usepackage{array}

\def\be{\begin{equation}}
\def\ee{\end{equation}}
\def\bea{\begin{eqnarray}}
\def\eea{\end{eqnarray}}
\def\ba{\begin{array}}
\def\ea{\end{array}}
\def\bdm{\begin{displaymath}}
\def\edm{\end{displaymath}}

\begin{document}



\title{
Coupling of acoustic phonon to a spin-orbit entangled pseudospin\\
}


\author{S.-K. Yip}
\affiliation{Institute of Physics, Academia Sinica, Taipei 115, Taiwan}
\affiliation{Institute of Atomic and Molecular Sciences, Academia Sinica, Taipei 106, Taiwan}



\date{\today}

\begin{abstract}

We consider coupling of acoustic phonon to pseudospins consisting of  electronic spins locked to orbital angular momentum states. 
We show that a Berry phase term arises from projection onto the time-dependent lowest energy manifold. 
We examine consequences on the phonon modes, in particular mode splitting,
 induced chirality and Berry curvatures under an external magnetic field which Zeeman couples to the pseudospin.  

\end{abstract}


\maketitle



\section{Introduction}

How phonons couple to magnetic field has received a lot of attention recently, with particular boost due to the interest in thermal  Hall effects
and the question of possible phonon contributions \cite{Strohm,Ideue,Kasahara18a,Kasahara18b,Grissonnanche19,Hentrich,Li20,Boulanger20,Lefrancois22,Sim21}. 
In this paper, we investigate a mechanism of phonon-magnetic field coupling thereby an acoustic phonon can acquire a Berry curvature, and the otherwise degenerate phonon modes (in the absence of this coupling) would be mixed, producing chiral modes with finite frequency splitting.    
The general mechanism of generating such coupling between the phonon to the magnetic field is by now well-appreciated.
While in the case of optical phonons in strongly ionic solids, the coupling can be directly comprehended as due to motion of the charged ions \cite{Juraschek19},  
in general it has to be understood as a Berry phase effect
 \cite{Mead,Qin12,Saito19,Saparov22,Bonini23}.  
 Phonons are associated with the motion of the atoms or ions in the solid. 
The electrons, on the other hand, not only provide an effective scalar potential between the ions given in the traditional Born-Oppenheimer approximation,
but also carries a Berry phase factor depending on the ionic coordinates. 
This phase factor, after the electron degrees of freedom have been eliminated, gives rise to
  an effective vector potential  \cite{Mead,Qin12,Saito19,Saparov22,Bonini23,Mead92} and hence Lorentz force for the motions of the ions or nuclei.
Traditional first principle phonon calculations in solids based on density functional theory \cite{LDA} take into account electron-phonon interactions only via the ``interatomic force constant" matrix
and thus miss the Berry phase contribution mentioned above, though more recent works (e.g. \cite{Bonini23}) have allowed for this contribution. 
The generation of gauge field on one subsystem via projecting out the other  has also been discussed in other branches of physics (e.g. \cite{Wilczek,Stone,Goldman}). 

We shall here consider phonons coupling to the magnetic field via spins.  We shall primarily consider localized spins in the paramagnetic regime, where the spins are not ordered or even non-interacting, with finite polarization  only due to the external applied magnetic field.
The coupling mechanism we consider is different from those investigated in the literature,
such as magnetic anisotropy energy \cite{Kittel58} in magnetically ordered systems, 
or modifications of spin-spin interaction energies due to bond-length or angle changes in the presence of phonons. 
The specific systems we shall examine are those where the ``spins" are actually pseudospins, with electronic spins entangled with orbital angular momentum states,
for examples,  
 Ru$^{+3}$ ions in $\alpha-$RuCl$_3$ \cite{Kasahara18a,Kasahara18b,Hentrich,Lefrancois22}, or  Ir$^{+4}$ in Sr$_2$IrO$_4$ \cite{Kim08,Kim09,Shitade09,JK09}
with (Kramers degenerate) ground states well separated from excited states \cite{eta}. 
Systems with such strong spin-orbit entangled pseudospins themselves are under strong recent attention due to interesting physics such
as spin-orbit assisted Mott transition,  unusual interaction between pseudospins, possible spin liquids and multipolar order etc. \cite{Takayama}. 
In the presence of the acoustic phonon, the local environment becomes time dependent. 
If the pseudospin is not excited, then this pseudospin must remain within the ground state manifold though defined according to this instantaneous environment. 
This time dependence then generates an effective gauge field for the ionic motion. 
 Since the pseudospin Zeeman couples to the magnetic field, 
direct phonon-magnetic field coupling would result, providing the mechanism we desire in the first paragraph. 
Explicitly we shall be examining d-electron systems in cubic environment.
However, the mechanism seems to be quite general when both crystal field splitting and strong spin-orbit coupling are present
when the  phonon frequencies lie within suitable frequency ranges. 
Since a projection into a subspace is necessary, our mechanism is only applicable for such strongly spin-orbit entangled systems. 

Our mechanism to be discussed here is distinct from the one which has been investigated also for spin-orbit entangled pseudospins 
in particular for f-electron systems (e.g. \cite{Thalmeier,Schaack,Capellmann89,Capellmann91}) coupling to optical phonons.
There, the coupling,  termed magneto-elastic interaction in \cite{Thalmeier,Schaack,Capellmann89,Capellmann91}
(but to be distinguished from magneto-elastic couplings which has been discussed in magnetostriction or for acoustic waves in , e.g., \cite{Kittel58,book}), 
arises from the modification of crystal fields acting on the pseudospins in the presence of the optical phonons. 
These phonon-pseudospin couplings are parameterized by coupling constants which describe thus the extent that the crystal fields are modified due to the
displacements of the ions surrounding the pseudospin under discussion.
In this mechanism, the splitting of degenerate phonon modes by the magnetic field is generated by virtual transitions between different energy manifolds \cite{Thalmeier,Schaack}. 
In contrast, our mechanism arises from phase factors generated from projection onto the time-dependent pseudospin ground state manifold.
As we shall see, the ``coupling constant" depends on the information entirely of the ground state manifold, and in fact 
a factor related to the geometric information on the structure of the pseudospin. 


The structure of the rest of this paper is as follows.  In Sec. \ref{sec:model} we introduce our specific model, and then derive
the phonon-pseudospin coupling.   The effect of this coupling on the sound modes frequencies is evaluated in Sec. \ref{sec:sound}.
In Sec.  \ref{sec:Berry} we evaluate the Berry curvatures.  We end with some order of magnitude estimates and discussions in Sec. \ref{sec:discussions}.

\section{Model}\label{sec:model}

To be specific, consider Ir$^{+4}$ ions Sr$_2$IrO$_4$ or Ru$^{+3}$ ions in RuCl$_3$, both with five $d$ electrons. (see, e.g. \cite{Kim08,Kim09,Shitade09,JK09})  In both cases, the ions are situated within an approximately cubic environment formed by the O$^{-2}$ and Cl$^{-1}$ ions,  respectively.  The $d$-electrons energy levels are crystal-field split into a $t_{2g}$ and an $e_{2g}$ manifold.   Only the $t_{2g}$ manifold consisting of the orbitals usually labelled as $xy$, $yz$, and $zx$ are relevent, and together with the electronic spin $\uparrow$ and $\downarrow$ degree of freedom, form six levels.   The spin-orbit interaction further splits these six levels into one quartet, usually labelled as $j_{eff} = 3/2$, which are occupied, and another
Kramer's doublet, usually labelled as $j_{eff} = 1/2$, which is singly occupied.   
 We shall write the wavefunctions for the two levels in this doublet as \cite{notei}
\bea
\vert \Uparrow \rangle &=  \frac{ -i}{\sqrt{3}} & \left[ \vert xy \uparrow \rangle + \vert yz \downarrow \rangle + i \vert xz \downarrow \rangle \right] \nonumber \\
\vert \Downarrow \rangle &=  \frac{ i}{\sqrt{3}}  &\left[ \vert xy \downarrow \rangle -  \vert yz \uparrow \rangle + i \vert xz \uparrow \rangle \right] \label{ps}  \ ,
\eea
forming a time-reversal pair (we use the convention, under time-reversal, $\vert \uparrow\rangle \to \vert \downarrow \rangle $, $\vert \downarrow \rangle  \to - \vert  \uparrow \rangle$,  and
similarly,  $\vert \Uparrow\rangle \to \vert \Downarrow \rangle $, $\vert \Downarrow \rangle  \to - \vert  \Uparrow \rangle$). 
In the absence of phonons, the orbital parts of the wavefunctions ($xy$, $yz$, $zx$) as well as the spin parts ($\uparrow$, $\downarrow$)
 are defined according to fixed axes with respect to the crystal in equilibrium.

Before we consider phonons, let's first note a few relations which we shall use.
Denoting the electronic spin operator by $\vec s = \frac{1}{2} \vec \sigma$ where $\vec \sigma$ are Pauli matrices operating on the $\uparrow$ and $\downarrow$ space,
and $\vec L$ the orbital angular momentum operator, their projections onto the subspace of eq.  (\ref{ps}) are  \cite{noteLs}
\bea
\vec s = - \frac{1}{6} \vec \tau  \ , \qquad 
\vec L = -\frac{2}{3} \vec \tau  \ , \label{Lstau}
\eea
where $\vec \tau$ are Pauli matrices within the within the  $\Uparrow$, $\Downarrow$ space. 
The energy change under a magnetic field $\vec B$,  $ \mu_B (\vec L + 2 \vec s) \cdot \vec B$ (with $\mu_B$ the Bohr magneton) 
with the operators projected again onto this subspace
({\it i.e.}, ignoring thus other contributions), would then be 
\be
 E_Z = 
\mu_B ( - \frac{2}{3} - \frac{1}{3} ) \vec \tau \cdot \vec B 
 \equiv - g \mu_B \frac{\vec \tau}{2} \cdot \vec B  \label{Zeeman}
\ee
with an effective $g$ fector of $2$ \cite{Kim09,Shitade09}. 
In the first equality of eq (\ref{Zeeman}), $-\frac{2}{3}$ arises from $\vec L$ and 
$-\frac{1}{3} = 2 \times ( - \frac{1}{6}) $ arises from $2 \vec s$. 
Eq (\ref{Lstau}) implies
\be
\vec L + \vec s = - \frac{5}{6} \tau \  ,
\label{J}
\ee
a result which we shall use later. 

\subsection{phonon-pseudospin coupling}

Consider a long wavelength acoustic phonon, with a spatial and time dependent displacement vector $\vec \xi (\vec x, t) $.  For simplicity,
we shall consider a cubic crystal, and remark on modifications for other symmetries later.   As is well-known,
we can decompose this into three components:  $\vec \nabla \cdot \vec \xi$, $\frac{1}{2} \vec \nabla \times \vec \xi$ and 
the tensor $\frac{1}{2} \left( \frac{\partial \xi_l} {\partial x_j} + \frac{\partial \xi_j} {\partial x_l} \right) - \frac{1}{3}  \delta_{jl} \vec \nabla \cdot \vec \xi$,
corresponding to an isotropic expansion (compression), rotation, and anisotropic deformation respectively \cite{LL}.
Under a low energy excitations of the crystal \cite{eta}, the electronic state $\vert \Psi \rangle $ of our ion under consideration should
still be within the manifold described by eq (\ref{ps}) though in a frame specified by the local environment.  
Hence at an instantaneous time $t$, we should have (up to small terms describing the excitations to higher energy levels)
\be
\vert \Psi (t) \rangle  = \alpha'_{\Uparrow} (t) \vert \Uparrow' (t) \rangle +  \alpha'_{\Downarrow}(t)  \vert \Downarrow' (t) \rangle 
\label{Psi}
\ee
where $\vert  \Uparrow' (t) \rangle$ ( $\vert \Downarrow' (t) \rangle$) are states given by eq (\ref{ps}) except with
$x, y, z$, $ \vert \uparrow \rangle, \vert \downarrow \rangle $ replaced by $x',y',z' $, $\vert \uparrow' \rangle,  \vert \downarrow' \rangle$
rotated from the former by $\vec \Theta(t)  \equiv \frac{1}{2} \vec \nabla \times \vec \xi(t)$. (The isotropic compression
and anisotropic deformation
 would not affect what we would be discussing below \cite{drop} and shall be ignored from now on). 
Suppose that our ion is under an external field $\vec B$, and let $\vec B'$ be the value of this field in the above mentioned rotating frame. 
The Schr\"odinger equation of motion for $\vert\Psi\rangle$, 
employing eq. (\ref{Psi}) and noting
the time dependence of the basis function $\vert \Uparrow' (t) \rangle, \vert \Downarrow' (t) \rangle$,  implies
\begin{widetext}
\be
i \frac{\partial} {\partial t} \left( \ba{c} \alpha'_{\Uparrow} \\ \alpha'_{\Downarrow} \ea \right)
 =    - g \mu_B \vec B'(t) \cdot \frac{\vec \tau}{2}  \left( \ba{c} \alpha'_{\Uparrow} \\ \alpha'_{\Downarrow} \ea \right) + 
\left( \ba{cc} 
  -i \langle \Uparrow' \vert \frac{\partial}{\partial t}  \vert \Uparrow' \rangle & 
  - i  \langle \Uparrow' \vert \frac{\partial}{\partial t}  \vert \Downarrow' \rangle \\
 -i \langle \Downarrow' \vert \frac{\partial}{\partial t}  \vert \Uparrow' \rangle  & 
 -i \langle \Downarrow' \vert \frac{\partial}{\partial t}  \vert \Downarrow' \rangle \ea \right) 
 \left( \ba{c} \alpha'_{\Uparrow} \\ \alpha'_{\Downarrow} \ea \right)
\ee
\end{widetext}
Here $\vec \tau$, which rigorously should have been denoted as $\vec \tau'$, are Pauli matrices in the
$\Uparrow'$, $\Downarrow'$ subspace,  but we shall not make this distinction in notations for simplicity.
Since $\vert \Uparrow'(t) \rangle =  e^{- i \vec \Theta \cdot ( \vec L + \vec s) } \vert \Uparrow \rangle$
$\approx   (1 - i \vec \Theta \cdot ( \vec L + \vec s)) \vert \Uparrow \rangle $, 
the time derivatives can be evaluated as , e.g., 
$  -i \langle \Uparrow' \vert \frac{\partial}{\partial t}  \vert \Uparrow' \rangle $ $ = 
 - ( \frac{\partial \vec \Theta}{ \partial t} ) \cdot  [ \langle \Uparrow' \vert  ( \vec L + \vec s)   \vert \Uparrow' \rangle ]  $.
Using eq (\ref{J}) (and ignoring a terms  $\propto \vec \Theta \times \frac{\partial \Theta}{ \partial t} $
 which arises due to the difference between the primed and unprimed $\Uparrow$ and $\Downarrow$ space), we obtain
\be
i \frac{\partial} {\partial t}
 \left(  \ba{c} \alpha'_{\Uparrow} \\ \alpha'_{\Downarrow} \ea \right)
=  \left[ - g \mu_B \vec B' (t)\cdot \frac{\vec \tau}{2} 
+ \frac{5}{6} \frac{\partial \vec \Theta}{\partial t} \cdot \vec \tau \right] 
\left( \ba{c} \alpha'_{\Uparrow} \\ \alpha'_{\Downarrow} \ea \right)
\ee
It would be more convenient to have an equation of motion involving directly $\vec B$  instead.
We observe that $\vec B' = \vec B  - \vec \Theta \times \vec B$ and hence
$ \vec B' \cdot \vec \tau =  $ $   e^{  i \frac{\vec \Theta}{2} \cdot \tau} \vec B  \cdot \vec \tau e^{ - i \frac{\vec \Theta}{2} \cdot \tau}$.
Introducing
\be
  \left(  \ba{c} \tilde \alpha_{\Uparrow} \\ \tilde \alpha_{\Downarrow} \ea \right) 
 = e^{ - i \frac{\vec \Theta}{2}  \cdot \tau }  \left(  \ba{c} \alpha'_{\Uparrow} \\ \alpha'_{\Downarrow} \ea \right)
\ee
we obtain finally
\be
i \frac{\partial} {\partial t}
 \left(  \ba{c} \tilde  \alpha_{\Uparrow} \\ \tilde \alpha_{\Downarrow} \ea \right)
= \left[ - \frac{g \mu_B}{2} \vec B   
+ \frac{4}{3} \frac{\partial \vec \Theta}{\partial t}\right]  \cdot \vec \tau  
\left( \ba{c} \tilde \alpha_{\Uparrow} \\  \tilde \alpha_{\Downarrow} \ea \right)
\ee
where we have again dropped a term involving second powers in $\Theta$ . 
$\frac{4}{3}$ arises from $  \frac{1}{2}  - ( -  \frac{5}{6})$ thus is due to the difference between
the rotational matrix for ordinary spin-1/2 
and our pseudospin (eq. (\ref{J})). 
The direction of the pseudospin, defined as the
expectation value of $\vec \tau$ with the ``spin" wavefunction $(\tilde \alpha_{\Uparrow}, \tilde \alpha_{\Downarrow})$,
is given by
\be
\frac{\partial}{\partial t} \hat \tau = \hat \tau \times \left[ \vec \omega_0  +r   \frac{\partial \vec \Theta}{\partial t}  \right] 
 = \hat \tau \times \left[ \vec \omega_0 + \frac{r}{2} ( \nabla \times \frac{\partial \vec \xi}{\partial t} ) \right]
\label{precess}
\ee
with $\vec \omega_0 = g \mu_B \vec B$ and $r = - \frac{8}{3}$.  The former is the standard precession due to the external field and the second
extra term is due to the rotational properties of our basis functions derived above.   

\subsection{Lagrangian} 

We construct now the Lagrangian for the coupled phonon and pseudospin system.  To simplify the writing,
when no confusion arises, we shall often just write ``spin" for the pseudospin. 

First, the acoustic phonon alone can be described by the Lagrangian density
\be
L_{0,ph} = \frac{1}{2} \rho_M \left( \frac{ \partial  \xi_j} { \partial t} \right)^2 - U_{elastic}  \label{Lph}
\ee
where 
$U_{elastic} = \frac{1}{2} \left[ \lambda_1 (\frac{ \partial \xi_j}{\partial x_l} \frac{ \partial \xi_j}{\partial x_l}) +
 \lambda_2 \frac{ \partial \xi_j}{\partial x_j} \frac{ \partial \xi_l}{\partial x_l} \right]$ is the elastic energy density. 
Here $\rho_M$ is the mass density (dimension mass times inverse volume)  and sums over repeated indices are implicit. 
We have also ignored a term 
$\lambda_3  (\frac{ \partial \xi_j}{\partial x_j} \frac{ \partial \xi_j}{\partial x_j}) $ which
is allowed in cubic symmetry for simplicity.  Its effects will be discussed later. 
Under this simplification, for a system without coupling to spin,  sound velocities are independent of 
direction of propagation $\hat q$, with longitudinal and transverse sound velocities given by
$v_L = [ (\lambda_1 + \lambda_2) / \rho_M]^{1/2} $ and $v_T = [\lambda_1/ \rho_M]^{1/2}$ respectively. 

For the spin, first we recall that, for a spin $S$ under a magnetic field along $\hat z$,  the Lagrangian can be
written as \cite{Altland}
 $ L_s = g \mu_B S B \cos \theta  + S \cos \theta \frac{\partial \phi}{\partial t} $
where $\theta$ and $\phi$ are the angles for the spin direction in spherical coordinates, the first term being from the Zeeman energy and
the second a Berry phase term. 
To produce  the equation of motion (\ref{precess}), we need only to replace 
$ g \mu_B S B \cos \theta$ by
$ \frac{\vec \tau}{2} \cdot \left[ g \mu_B  \vec B + \frac{r}{2} ( \nabla \times \frac{\partial \vec \xi}{\partial t} ) \right] $
(now specializing to pseudospin $1/2$). The last term allows us to identify the pseudospin - phonon coupling. 

The Lagrangian $L = L_{ph} + L_{s} + L_{ph-s}$ is a sum of the phonon term (\ref{Lph}), the spin
term and the phonon-spin coupling term. We then have, for a net effective spin density $\rho_s$ per unit volume, 

\be
L_s = \rho_s \frac{1}{2}  \left[ g \mu_B \vec B \cdot \hat \tau + \rm cos \theta  \frac{\partial \phi}{\partial t} \right]  \label{Ls}
\ee

\be
L_{ph-s}  = \frac{r \rho_s}{4} \left[ \hat \tau \cdot  ( \nabla \times \frac{\partial \vec \xi}{\partial t} ) \right] \label{Lphs}
\ee
with $\hat \tau$ the net (pseudo-)spin direction.  The phonon-pseudospin coupling is dictated by the factor $r$ derived in the last subsection. 
As is evident from our derivation above, this coupling arises from the Berry phase due to the rotating frame of reference for the pseudospin in the presence
of the transverse acoustic phonon. 
We remind the readers here that this coupling thus has an entirely different origin from the magneto-elastic coupling discussed by, e.g., \cite{Kittel58}
for magnetic materials, which describes the change in magnetic energies  in the presence of stress. 


\subsection{ effective equation of motion}

The equation of motion for $\hat \tau$ was already obtained in (\ref{precess}), which reads, after
Fourier transform and linearizing about the equilibrium where $\hat \tau = \hat z$, 

\be
- i \omega \hat \tau (\omega, \vec q) = \omega_0 ( \hat \tau \times \hat z) + \frac{ r \omega}{2} \left[ \hat z \times ( \vec q \times \vec \xi) \right] 
\label{spin}
\ee
where $\vec q$ is the wavevector and $\omega$ the angular frequency. 

The equation of
motion for the displacement is

\be
\rho_M \omega^2  \xi_j - \frac{r \omega}{4} \rho_s ( \vec q \times \hat \tau)_j =
\lambda_1 q^2 \xi_j + \lambda_2 q_l ( q_j \xi_l) 
\label{xi}
\ee

We now study the consequences of eq. (\ref{spin}) and (\ref{xi}).  Equation (\ref{spin}) implies
that $\hat \tau_z$ is just a constant.   The components orthogonal to the field direction  ($j = x, y$) obeys


\be
\tau_j   = \frac{ r \omega / 2}{ \omega_0^2 - \omega^2 } \left[
 \omega_0 ( \vec q \times \vec \xi )_j -  i \omega  (\hat z \times (\vec q \times \vec \xi))_j \right]
\label{tauxy} 
\ee

Putting this into eq. (\ref{xi}) gives us the equation of motion entirely expressed in terms of $\xi_j$:
\begin{widetext}
\be
0 =  \rho_M \omega^2  \xi_j - \left[ \lambda_1 q^2 \xi_j + \lambda_2 q_l ( q_j \xi_l)  \right]
- \frac{ r^2 \rho_s}{8} \frac{ \omega_0 \omega^2 }{ \omega_0^2 - \omega^2} 
\left[ - q_z^2  \xi_j + q_z q_j \xi_z + ( q_z ( q_l \xi_l) - q^2 \xi_z) \delta_{jz}  \right]
- i  \frac{ r^2 \rho_s}{8}  \frac{\omega^3 }{ \omega_0^2 - \omega^2} q_z  (\vec q \times \vec \xi)_j
\label{xieff}
\ee
\end{widetext} 
Coupling of  the pseudospin to the phonon results in the last two new terms. 
Here the factor  $\delta_{jz} = 1$ if $j=z$ and vanishes otherwise. 
We note the factor $q_z$ in the last term, which is generated from the last term in eq. (\ref{tauxy}).  This factor
reflects the fact that the time dependent parts of $\tau$ only have $x$ and $ y$ components. 

We now analyze eq. (\ref{xieff}) in two different limits.

\section{sound modes}\label{sec:sound}

\subsection{small magnetic field: anti-adiabatic regime}\label{sound-anti}

For small fields, $\omega_0$ is much smaller than the phonon frequencies, eq. (\ref{xieff}) approximately
reads
\be
0 =  \rho_M \omega^2  \xi_j -  \left[ \lambda_1 q^2 \xi_j + \lambda_2 q_l ( q_j \xi_l)  \right]
+  i  \frac{ r^2 \rho_s}{8}  \omega q_z  (\vec q \times \vec \xi)_j
\label{xieff-h}
\ee

Longitudinal sound, with $\xi$ parrallel to $\vec q$,  is not affected.  
 Physically, there is no rotation of the environment surrounding the pseudospin in this case.
The two polarizations of the transverse sound are coupled via
the spins, turning them into circular polarized ones.  Writing $\vec \xi = \xi_{\theta} \hat \theta + \xi_{\phi} \hat \phi$,
we get
\be
\left( \ba{cc} \omega^2 - q^2 v_T^2   &  - i \frac{\rho_s r^2}{ 8 \rho_M} \omega q^2 \cos \theta_q \\
 + i \frac{\rho_s r^2}{ 8 \rho_M} \omega q^2 \cos \theta_q  & \omega^2 - q^2 v_T^2 \ea  \right) 
\left( \ba{c} \xi_\theta \\ \xi_{\phi} \ea \right) = 0
\label{xim-h}
\ee
Here $\theta_q$ is the angle between $\hat q$ and $\hat z$.
  To lowest order in the phonon-pseudospin coupling, the frequencies are  given by 
\be
\omega_{\pm} = q v_T \left[ 1 \pm Z \rm cos \theta_q \right]
\label{omega-h}
\ee
for the modes with right ( $ (\xi_{\theta}, \xi_{\phi} ) \propto ( 1, i) $) and left 
( $ (\xi_{\theta}, \xi_{\phi} ) \propto ( 1, - i) $) circular polarization, 
 with $Z$  a $q$-dependent dimensionless parameter
\be
Z \equiv \frac{ \rho_s r^2 q}{ 16 \rho_M v_T}  \ .
\label{Z}
\ee
Thus the fractional splitting increases with $q$, reflecting that a shorter wavelength implies a larger rotation motion
of the lattice $\vec q \times \vec \xi$ and hence a stronger coupling to our pseudospin. 
This is different from a na\"ive picture of hybridization between the phonon modes with the Larmor precession of the spins,
where the induced splitting would decrease with increasing frequencies away from  $\omega_0$.  
From eq. (\ref{omega-h}), we see that
for $q_z > 0$, the lower (higher) frequency mode is left (right)-circularly polarized.
The reverse is the case if $q_z < 0$.  See Fig \ref{fig1}a. 

\begin{figure*}[tbp]\label{fig1}
\begin{center}
\includegraphics[width=0.9\columnwidth]{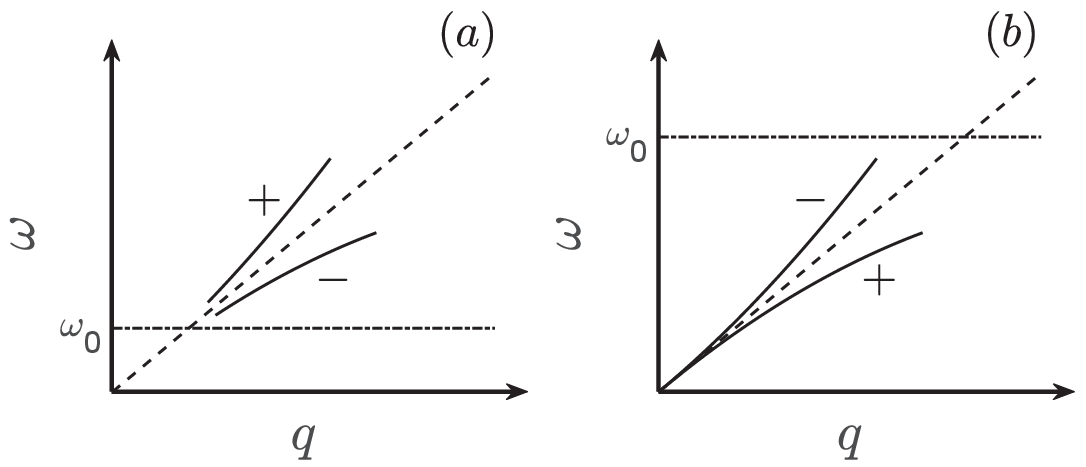}
\caption{
Schematic dispersions for the transverse phonon modes for $q_z > 0$. $+$ ($-$) labels right (left) circularly or elliptically polarized.  For $q_z < 0$,
the $\pm$ labels in the above figures have to be reversed. 
}
\label{fig1}
\end{center}
\end{figure*}

\subsection{low frequency:  adiabatic regime}\label{sound-low}

For very small $q$, the phonon frequency $\sim  q v_T$ is much smaller than $\omega_0$.
In this case, the effective equation of motion for the phonon cooridinate can be written as
\begin{widetext}
\be
0 =  \rho_M \omega^2  \xi_j - \left[ \lambda_1 q^2 \xi_j + \lambda_2 q_l ( q_j \xi_l)  \right]
-  \frac{ r^2 \rho_s \omega^2}{8  \omega_0  }
\left[ - q_z^2  \xi_j + q_z q_j \xi_z + ( q_z ( q_l \xi_l) - q^2 \xi_z ) \delta_{jz} \right]
 -  i  \frac{ r^2 \rho_s}{8}  \frac{\omega^3 }{ \omega_0^2} q_z  (\vec q \times \vec \xi)_j
\label{xieff-l}
\ee
\end{widetext} 

Note the sign differences between the last terms of eqs. (\ref{xieff-h}) and (\ref{xieff-l}) in 
two different frequency regimes, similar to the case of, e.g., driven harmonic oscillator for above versus below resonance. 
Formally the last term is one higher order in $\omega_0^{-1}$ than the second last, but we shall explain shortly why we keep this term. 
Longitudinal sound is again unaffected.  The eigenvector has $\vec \xi$ parallel to $\vec q$,
as can be checked by multiplying eq. (\ref{xieff-l}) by $q_j$ and the sum over $j$
(there is no contribution from either the last or second last terms).   
The transverse sounds obey
\begin{widetext}
\be
\left( \ba{cc} \omega^2 - q^2 v_T^2  +   \frac{\rho_s r^2} { 8 \rho_M \omega_0} q^2  \omega^2  
&   i \frac{\rho_s r^2}{ 8 \rho_M \omega_0^2} \omega^3 q^2 \cos \theta_q \\
 -  i \frac{\rho_s r^2}{ 8 \rho_M \omega_0^2} \omega^3 q^2 \cos \theta_q  &
 \omega^2 - q^2 v_T^2  +  \frac{\rho_s r^2}{ 8 \rho_M \omega_0} q^2 \omega^2  \cos^2 \theta_q \ea  \right) 
\left( \ba{c} \xi_\theta \\ \xi_{\phi} \ea \right) = 0
\ee
\label{xim-l}
\end{widetext}

For $\theta_q$ not too close to $0$ or $\pi$, we can ignore the off-diagonal terms in this matrix equation
as they are second order in $\omega_0^{-1}$. We obtain two non-degenerate modes
with frequencies
$\omega = q v_T  (1 + X)^{-1/2}$ (for $\vec \xi$ along $\hat \theta$) and $\omega = q v_T / (1 + X \cos^2 \theta_q )^{-1/2}$ (for $\vec \xi$ along $\hat \phi$).
Here $X \equiv \frac{\rho_s r^2 q^2 }{ 8 \rho_M \omega_0} $ is a q-dependent  dimensionless parameter. 
Thus the mode with $\vec \xi$ along $\hat  \theta$ has a lower frequency than the one with $\hat \phi$ due to the coupling to the pseudospin.  
For $\theta_q = 0$ or $\pi$, these two modes are degenerate up to $\omega_0^{-1}$.  The off-diagonal term
then turns these transverse modes to circularly polarized.
For $\theta_q = 0$,  the modes with $ (\xi_{\theta}, \xi_{\phi} ) \propto ( 1, \pm i) $ have frequencies roughly given by
$ \omega \approx  q v_T  (1 + X)^{-1/2} [ 1 \mp X_2 ] $, with the dimensionless parameter
$X_2  \equiv \frac{\rho_s r^2 q^2 }{ 16 \rho_M \omega_0} \frac{q v_T} {\omega_0}$.
Note that both $X$ and $X_2$ are increasing functions of $q$. 
Similar to the case in subsection \ref{sound-anti},  the sign in front of $X_2$ in this expression for $\omega$ needs to be reversed for $\theta_q = \pi$.
Note that $X_2 \ll X$ since we are now considering $q v_T \ll \omega_0$ and also that
the circular polarization for the higher frequency mode is opposite to the anti-adiabatic case for a given $\hat q$. 
For general $\theta$, the modes are elliptically polarized. 
See Fig \ref{fig1}b. 

\section{Berry curvature}\label{sec:Berry}

We here discuss the Berry curvature for the phonon modes.   Our methodology here closely follows  \cite{Matsumoto}
 and Supplemental Materials of \cite{CKS}.
In the Appendix we collect some of the relevant formulas.  
We shall again first investigate the small magnetic field regime (Sec. \ref{Berry-h}) and then the high magnetic one (Sec. \ref{Berry-l})
The second regime is included here for completeness but the information therein is not essential for our final Discussion section,
so readers can choose to skip Sec. \ref{Berry-l}.

\subsection{Anti-abiabatic}\label{Berry-h}

The Lagrangian density that reproduces the equation of motion  (\ref{xieff-h}) can easily found to be
\be
L = L_{0,ph}  
      + \frac{ r^2 \rho_s}{16} \epsilon_{jkl}   \left( \frac{\partial^2 \xi_j}{\partial z  \partial x_k} \right) \left( \frac{\partial \xi_l }{\partial t} \right)
\label{L-anti}
\ee
The last term, in the form of an effective Lorentz force,  might have been expected from phenomenological grounds.  An initial guess might be a term proportional to
$\hat z \cdot (\vec \xi \times \frac{\vec \partial \xi}{\partial t})$:  this term  does arise in the case of optical phonons
\cite{Capellmann89,Capellmann91,Juraschek22},  but here this is not acceptable since 
the appearance of $\vec \xi$  violates translational invariance.  Instead, in eq. (\ref{L-anti}),  a second order spatial derivative
appears instead, similar to what has been discussed in \cite{Qin12,CKS},
though in our case the precise form, as derived in Sec \ref{sec:model},  is different here. 

The conjugate momentum $\Pi_j$ is given by
\be
\Pi_j  \equiv \frac{\partial L}{ \partial \dot \xi_j}  = 
\rho_M  \left( \frac{\partial \xi_j}{\partial t} \right) - \frac{ r^2 \rho_s}{16} \epsilon_{jkl}   \left( \frac{\partial^2 \xi_l }{\partial z \partial x_k} \right)
\label{Pi1}
\ee
with the equation of motion (\ref{xieff-h}) just the same as
$\frac{ \partial \Pi_j}{ \partial t} = \frac{\partial L}{ \partial \xi_j}$. 
After Fourier transforming the spatial coordinates,  these two equations can be written in matrix form

\be
\frac{\partial}{\partial t} \left( \ba{cc} \rho_M \hat 1 & 0 \\ \rho_M \Omega & \hat 1 \ea \right) 
\left( \ba{c} \xi \\ \Pi \ea \right) 
= \left( \ba{cc} - \rho_M \Omega & \hat 1 \\ - \mathcal{ Q} & 0 \ea \right) \left( \ba{c} \xi \\ \Pi \ea \right) 
\label{matrix1}
\ee
where $\Omega$, $\mathcal{Q}$,  $\hat 1$ are $3 \times 3$ matrices: 
 $\Omega \equiv
Z (q v_T) \cos \theta_q  \hat \Omega$ with $Z$ defined in eq (\ref{Z}), 
$\hat \Omega_{jk} \equiv - \epsilon_{jkl} \hat q_l$, 
$\mathcal{Q}_{jk} \equiv \lambda_1 q^2 \delta_{jk} + \lambda_2 q_j q_k$,
and $\hat 1_{jk} = \delta_{jk}$. 

Eq (\ref{matrix1}) can be rewritten as
\be
\frac{\partial}{\partial t} \left( \ba{c} \xi \\ \Pi \ea \right)  = 
-  i \mathcal{S}  \left( \ba{c} \xi \\ \Pi \ea \right) 
\label{matrix2}
\ee
with $\xi$, $\Pi$ column matrices consisting of elements $\xi_{x,y,z}$ and $\Pi_{x,y,z}$, and  $\mathcal{S}$ a $6 \times 6$ matrix given by
\be
\mathcal{S} =  \left( \ba{cc} - i \Omega &  i / \rho_M  \\ - i  \mathcal{Q} & - i \Omega \ea \right)
\ee
where, rigorously speaking, the lower left element should have been $ - i \mathcal{Q} +  i  \rho_M  \Omega^2 $, and
we have taken the simpler form since $\Omega^2$ is second order in $1/\omega_0$ and hence higher order than the other terms we kept.  

Following \cite{CKS}, we search for the row vectors $(\vec u, \vec v)$ which satisfy, for positive frequencies $\omega$, 
\be
\omega (\vec u, \vec v) = (\vec u, \vec v) \mathcal{S}
\label{uv}
\ee
Once $(\vec u, \vec v)$'s are found, the Berry curvatures $\vec \Omega_B$ can then be evaluated
via  the formulas collected in Appendix \ref{curvature}. For the longitudinal mode, $(\vec u, \vec v) = (u_q \hat q, v_q \hat q)$.
The transverse modes can be more easily written in terms of $u_{\theta,\phi}$ and $v_{\theta, \phi}$
 defined via $\vec u = u_{\theta} \hat \theta + u_{\phi} \hat \phi$ and similarly for $\vec v$. 
They obey (observe that $\hat \theta \hat \Omega = - \hat \phi$ and $\hat \phi \hat \Omega = \hat \theta$)
\begin{widetext}
\be
\omega  \left( \ba{c} u_\theta \\  u_\phi \\ v_\theta \\ v_\phi \ea \right)
= \left( \ba{cccc}  & - i Z q v_T \cos \theta_q & - i \lambda_1 q^2 &  \\
            +  i Z q v_T \cos \theta_q & & & - i \lambda_1 q^2  \\
   i  \rho_M & & & - i Z q v_T \cos \theta_q \\
    &    i \rho_M   &  i Z q v_T \cos \theta_q  & \ea \right)
 \left( \ba{c} u_\theta \\  u_\phi \\ v_\theta \\ v_\phi \ea \right)
\ee
The right (left)  circular polarized mode has eigenvector  (normalized according to eq. (\ref{norm}))
\be
\left( \frac{ (\rho_M q v_T)^{1/2}}{2},   \pm \frac{ i (\rho_M q v_T)^{1/2}}{2},  \frac{i}{2 (\rho_M q v_T)^{1/2}},  \mp \frac{1}{2 (\rho_M q v_T)^{1/2}} \right)  ,
\label{eigen-h}
\ee
\end{widetext}
frequencies $\omega = q v_T ( 1 \pm  Z \cos \theta_q)$ (c.f. eq ( \ref{omega-h})) and curvature $\vec \Omega_B =  \pm \hat q / q^2 $.

\subsection{adiabatic}\label{Berry-l}

In this regime, eq. ({\ref{xieff-l}) indicates that the equation for the frequency is cubic.
This creates complications if we want to treat the problem in the same way as in the last subsection.   
However, since we are treating the pseudospin-phonon coupling as small, we
can simplify the problem by noting the fact that since the last term in eq. (\ref{xieff-l}) is thus already small, we can
replace $\omega^2$ there by the `` unperturbed" transverse sound frequency $ (q v_T)^2$ (transverse since
the last term affects only the transverse modes).  Thus we now consider the effective
equation of motion

\begin{widetext}
\be
0 =  \rho_M \omega^2  \xi_j - \left[ \lambda_1 q^2 \xi_j + \lambda_2 q_l ( q_j \xi_l)  \right]
-  \frac{ r^2 \rho_s \omega^2}{8  \omega_0  }
\left[ - q_z^2  \xi_j + q_z q_j \xi_z + ( q_z ( q_l \xi_l) - q^2 \xi_z ) \delta_{jz} \right]
-  i  \frac{ r^2 \rho_s}{8}  \frac{\omega (q v_T)^2 }{ \omega_0^2} q_z  (\vec q \times \vec \xi)_j
\label{xieff-l2}
\ee

This equation  reproduces the sound velocites discussed near the end of Sec. \ref{sound-low}
and we can check that the displacement eigenvectors found below are proportional to those found there. 


The Lagrangian density that reproduces this equation of motion can easily found to be
\be
L = L_{0,ph}  + \frac{r^2 \rho_s}{8 \omega_0}
 \left[ \frac{1}{2} \left( \frac{\partial^2 \xi_l }{\partial z \partial t} \right)^2 - \left( \frac{\partial^2 \xi_z}{\partial z \partial t} \right) \left( \frac{\partial^2 \xi_l}{\partial x_l  \partial t} \right)
  + \frac{1}{2} \left( \frac{\partial^2 \xi_z}{\partial x_l \partial t} \right)^2  \right]
      + \frac{ r^2 \rho_s v_T^2}{16 \omega_0^2} \nabla^2 \vec \xi \cdot  \vec \nabla \times  \left( \frac{\partial^2 \vec \xi}{\partial z \partial t} \right) 
\label{L-ad}
\ee


Carrying out the same procedure as in the last subsection, we obtain
\be
\frac{\partial}{\partial t} \left( \ba{cc} \rho_M ( 1 + X \hat \Lambda)  & 0 \\  - \rho_M \tilde \Omega & 1 \ea \right) 
\left( \ba{c} \xi \\ \Pi \ea \right) 
= \left( \ba{cc} \rho_M  \tilde \Omega & 1 \\ -\mathcal{Q}  & 0 \ea \right) \left( \ba{c} \xi \\ \Pi \ea \right) 
\ee
where 
 $\tilde \Omega \equiv X_2  q v_T \cos \theta_q \hat \Omega$ (dimension frequency)
with   $X, X_2$ defined in \ref{sound-low} and  $\hat \Omega_{jk}$ 
$\mathcal{Q}_{jk}$ already defined in subsection \ref{Berry-h},   
\be
\hat \Lambda \equiv \left( \ba{ccc}
\hat q_z^2 & 0 & - \hat q_x \hat q_z \\
0 & \hat q_z^2 & - \hat q_y \hat q_z \\
- \hat q_z \hat q_x & - \hat q_z \hat q_y & q_x^2 + q_y^2 \ea \right)
\ee

\end{widetext}

We have again the equation (\ref{matrix2}) with now
\be
\mathcal{S} =  \left( \ba{cc}  i [1 + X \hat \Lambda]^{-1} \tilde \Omega &  i / \rho_M [1 + X \hat \Lambda]^{-1}  \\
   - i  \mathcal{Q}  + i \rho_M \tilde \Omega [1 + X \hat \Lambda]^{-1} \tilde \Omega   &    i \tilde \Omega [1 + X \hat \Lambda]^{-1} \ea \right)
\ee
which, in accordance with our approximations,  the second term in the lower left element can be dropped. 

We can solve for the eigenvectors $(\vec u, \vec v)$ as before.  It is useful to note
the vector relations $\hat q \hat \Lambda = 0$, $\hat \theta \hat \Lambda = \hat \theta$ and $\hat \phi \hat \Lambda = \cos^2 \theta_q \hat \phi$. 
Once more, for longitudinal modes, $(\vec u, \vec v) = ( u_q \hat q, v_q \hat q)$ is unaffected by the pseudospin.
If $\theta_q$ is not too close to $0$ or $\pi$, in the first approximation we can ignore the effects of $\tilde \Omega$.  
The  modes are thus linearly polarized with either $\vec u$, $\vec v$ entirely along $\hat \theta$ or $\hat \phi$
with frequencies already given in subsection \ref{sound-low}. The 
 normalized eigenvectors are, respectively,
\begin{widetext}
\be
( u_\theta, v_\theta)_0 =  \left( \frac{ (\rho_M q v_T)^{1/2} ( 1 + X)^{1/4} }{\sqrt 2},    \frac{i}{\sqrt{2} (\rho_M q v_T)^{1/2} (1 + X)^{1/4} } \right) 
\label{thetam}
\ee
and 
\be
( u_\phi, v_\phi)_0 =  \left( \frac{ (\rho_M q v_T)^{1/2}( 1 + X \cos^2 \theta_q )^{1/4}  }{\sqrt 2},    \frac{i}{\sqrt{2} (\rho_M q v_T)^{1/2} ( 1 + X \cos^2 \theta_q)^{1/4}} \right) 
\label{phim}
\ee
for the lower and higher frequency mode. Here the subscript $0$ reminds us that we have ignored $\tilde \Omega$. 
The effect of finite $\tilde \Omega$  can be included by perturbation theory, using eqs. (\ref{thetam}) and (\ref{phim}) as the
``unperturbed" solutions.  
For the lower frequency mode, the wavevector can be written as  $ (\vec u, \vec v) = 
(u_{\theta, 0} \hat \theta, v_{\theta, 0} \hat \theta) + \beta ( u_{\phi, 0} \hat \phi, v_{\phi, 0} \hat \phi) $
where $\beta$ is  a small coefficient.   We find that $\beta$ is imaginary with 
\be
{\rm Im} \beta = \frac{X_2}{ 2 X} \frac{ \cos \theta_q}{ \sin^2 \theta_q} 
( 1 + X)^{1/4}   ( 1 + X \cos^2 \theta_q )^{1/4} \left[( 1 + X \cos^2 \theta_q )^{1/2}+  ( 1 + X \cos^2 \theta_q )^{1/2} \right]
\label{Imb}
\ee
hence ${\rm Im} \beta $ has the same sign as  $\cos \theta_q$.  For $q_z > 0$,    the lower frequency mode is right elliptically polarized (vice versa for $q_z < 0$). 
Similarly, the higher frequency mode (the $\phi$ mode before perturbation) becomes left elliptically polarized, with the degree of ellipticity 
characterized by the same coefficient ${\rm Im} \beta$.

For $\theta_q = 0$, the modes are circularly polarized, with normalized eigenvectors
\be
( u_\theta,  u_\phi, v_\theta, v_\phi) =  \left( \frac{ (\rho_M q v_T)^{1/2} ( 1 + X)^{1/4} }{2},  \mp \frac{ i (\rho_M q v_T)^{1/2} ( 1 + X)^{1/4} }{2}, 
  \frac{i}{2 (\rho_M q v_T)^{1/2} (1 + X)^{1/4} },  \frac{\pm 1}{2 (\rho_M q v_T)^{1/2} (1 + X)^{1/4} }  \right) 
\label{theta0}
\ee
for the higher  (left-circularly polarized) and lower (right-circularly polarized) frequency modes, respectively.
The opposite signs are to be taken if $\theta_q = \pi$. 

\end{widetext}

Eq. (\ref{Imb}) together with (\ref{thetam}) and (\ref{phim}) allow us to obtain the Berry curvature.
$\vec \Omega_B$ has no $\phi$ component. 
For $\theta_q$ not too close to $0$ or $\pi$, for the lower frequency mode, 
\be
\vec \Omega_B \cdot \hat \theta =  \frac{ 2 v_T}{q \omega_0} \frac{\cos^2 \theta_q}{\sin^3 \theta_q}   \ , 
\ee
\be
\vec \Omega_B \cdot \hat q =  \frac{4 v_T}{q \omega_0} \frac{\cos \theta_q}{\sin^4 \theta_q}  \ .
\ee
Here we have only kept the lowest order finite terms and have used $\frac{1}{q^2} \frac{ X_2}{X} = \frac{v_T}{2 q \omega_0}$.
For the higher frequency mode, there is an extra negative sign for these formulas. 

For $\theta_q = 0$, we obtain $\vec \Omega_B = \mp 1 / q^2$ for the two modes in eq. (\ref{theta0})
\cite{expand}.

\section{Discussions} \label{sec:discussions}

We begin with a rough estimate for the factor $Z$ in eq. (\ref{Z}),  which gives the fractional splitting in section  \ref{sound-anti}. 
Consider the case of one ion per unit cell, and 
let $\rho_0$ (dimension inverse volume) be the number of ions per unit volume, and $M$ is the mass per unit cell.  Then
$Z \approx \frac{\rho_s}{\rho_0} \frac{\hbar q}{ M v_T}$. 
( From here on we restore the Boltzmann constant $k_B$ and Planck constant $\hbar$.)
Suppose that $v_T \approx 1 {\rm km/sec}$, $M \sim  100$ proton mass,  
and if the spins are polarized ($\rho_s = \rho_0$), we get $Z \sim 10^{-3}$ for a $1$ meV phonon,
a very large value compared with those predicted in the literature \cite{Juraschek19,Bonini23} for other systems. 
For a paramagnet with small fields, $\rho_s / \rho_0 \sim \mu_B B /  k_BT$,
this number will be reduced, but 
still not necessarily small  for not too small fields and not too high temperatures. 

For the parameter $X$ in  sec \ref{sound-low}, ( note that $X \sim \frac{q v_T}{\omega_0} Z$) we obtain
$ X \approx  10^{-2}  \frac{\rho_s}{\rho_0} \frac{(\hbar q v_T/\rm meV)^2  }{  ( B /{\rm Tesla}) }  $.
For a $100$ Tesla field and $1$ meV phonon
we  have a $10^{-4}$ splitting if we take $\rho_s = \rho_0$. 

Phonons with finite Berry curvature will have an intrinsic  contribution  to the thermal Hall effect.
Though this contribution is seemingly small and unlikely to be at least the sole mechanism for the observed
thermal Hall effect for any systems, with thus extrinsic effects also called for (e.g. \cite{CKS, Mori}),
we here provide an estimate since it is often also evaluated
in the theoretical literature.  Considering small external magnetic field and the simplified situation in Sec. \ref{sound-anti} where we 
have two opposite circularly polarized modes, from the formulas in \cite{Qin12,Matsumoto} we estimate \cite{estimate}
$\kappa_{xy}/ T  \sim  \frac{\delta \omega}{v_T} \frac{k_B^2}{\hbar}$, where $\delta \omega$ is the typical splitting between 
the two oppositely polarized phonons at a given temperature , {\it i.e.}, 
  $\delta \omega \sim Z (q v_T)$ with $\hbar q v_T \sim k_B T$,   thus
\bdm
\frac{\kappa_{xy}}{T}  \sim  \frac{ \rho_s}{\rho_0} \frac{(k_B T)^2}{ \hbar  M v_T^3} \frac{ k_B^2}{\hbar}  \ .
\edm 
We obtain that $\kappa_{xy} > 0$ (see remark below eq (\ref{Z}) and footnote \cite{estimate}), 
independent of sign of $r$. 
Inserting the numbers, and taking again $\rho_s / \rho_0 \sim \mu_B B / k_B T$, we get
\be
 \kappa_{xy} \sim 10^{-8} ( T / {\rm K})^2 ( B / {\rm Tesla}) {\rm W / K m}  \ .  \label{estimate}
\ee
$\kappa_{xy}$ is proportional to $T^2$ instead of $T^3$ in \cite{Qin12,CKS} due
to the temperature dependence of $\rho_s$ just mentioned above. 
Eq. (\ref{estimate}) gives, for $B \sim 10$ Tesla and $T \sim 100$ K,  $\kappa_{xy} \sim$ mW/ K m,
a value comparable to those in, e.g., \cite{CKS},
and for $T \sim 30$ K, $\kappa_{xy} \sim 0.1$ mW/Km,  about an order of magnitude
smaller than the peak value found experimentally for the non-monotonic temperature dependent $\kappa_{xy}$
reported in \cite{Hentrich}.
Our number here however is likely to be an overestimate.  The Berry curvature
in our model relies on mixing between transverse modes.   If we take 
into account that rotational symmetries in crystals are discrete rather than
continuous, transverse phonon modes are already split for most propagating directions.
For these directions the sound modes are only ellipticallly polarized rather than circular, and the Berry curvature will be reduced.
A calculation would be similar to what we had in Sec.  \ref{Berry-l}.
Since the mixing term between the two transverse modes is $\sim Z q v_T$,
if the transverse mode velocites differ by $\Delta v_T$, the curvature
would be reduced by a factor $ \sim  Z / (\Delta v_T / v_T)$.

The mechanism discussed in this paper should be quite general, applicable to other systems so long
as the pseudospin has spin and orbital degrees of freedom entangled \cite{Takayama}
 with the lowest
multiplets not fully filled and not an orbital singlet, with energy well separated from the higher energy ones, when the phonon frequencies
lie within the suitable interval between these ``gaps".  Details will differ according
to the precise symmetry, and simple vector relation eq. (\ref{J}) between the rotational matrix and
the pseudospin Pauli matrices may not hold for lower symmetries, the proportionality 
factor $r$ will differ from our value given etc., but otherwise
the induced phase factors, mixing between phonon branches, and effective Lorentz forces will remain. 

Our mechanism would also be relevant for magnetically ordered systems.  In this case,
the coupling between the pseudospins that have been ignored so far will have to be taken into 
account, and our phonon-pseudospin coupling would appear as a phonon-magnon coupling.
There are already quite a number of papers dealing with phonon-magnon couplings 
\cite{Takahashi,Delugas} with interesting predictions,
furthermore mechanisms of inducing Berry curvature and chirality in the coupled phonon-magnon modes
have also been proposed (e.g. \cite{Takahashi}). 
However, our mechanism is of a qualitatively different nature
as it arises from the Berry phase generated from a time-dependent frame of reference 
of the pseudospin due to the sound mode. 
Instead, the mechanisms in \cite{Takahashi,Delugas} 
 ultimately are both based on the modifications of the spin-spin interactions due to the phonons,
with spin-orbital coupling arising from dipole-dipole interactions or magnetic anisotropy energies. 
(see also other theoretical works \cite{Ye20,Li23} for $\alpha-$RuCl$_3$).   To what extent our present mechanism will be important 
for magnetically ordered systems remains to be investigated.

\section{Acknowledgement}

This work is supported by the Ministry of Science and Technology,
Taiwan, under grant number MOST-104-2112-M-001-006-MY3.

\newcounter{seq}

\newenvironment{seq}{\refstepcounter{seq}\equation}{\tag{B\theseq}\endequation}

\appendix
\begin{center}
{\bf APPENDIX}
\end{center} 

Here we summarize some of the equations from \cite{Matsumoto} (hereafter MSM) and the Supplemental Materials of \cite{CKS} (CKS-SM)
that we have used in text. 
To simplify our notations, we shall drop labels corresponding to the components, different eigenvalues,  etc. 

\section{Eigenvectors }

After Fourier transform into wavevector $\vec q$ space, $\xi_{\vec q}$ and $\Pi_{\vec q}^{\dagger} = \Pi_{-\vec q}$ satisfies
the communtation relation
\be
[ \xi_{\vec q} , \Pi_{\vec q}^{\dagger} ]  = i \hbar 
\ee
Hence
\bea
\beta_{\vec q} &=  \frac{1}{\sqrt{2}} ( \xi_{\vec q} + i \Pi_{\vec q} ) \nonumber \\
\beta_{-\vec q}^{\dagger}  &=  \frac{1}{\sqrt{2}} ( \xi_{\vec q} - i \Pi_{\vec q} )
\eea
defines a set of annilhilation and creation operators. 
Let $\gamma_{\vec q}$, $\gamma_{-\vec q}^{\dagger}$ be instead the operators that
actually diagonalize the bosonic Hamiltonian, and define the transformation matrix between $\gamma_{\vec q}$ and $\beta_{\vec q}$
be $\mathcal{T}^{-1}$,  ({\it c.f.} MSM (6))), {\it i.e.}
\be
\left( \ba{c} \gamma_{\vec q} \\ \gamma_{-\vec q}^{\dagger} \ea \right) 
= \mathcal{T}^{-1} \left( \ba{c} \beta_{\vec q} \\  \beta _{-\vec q}^{\dagger} \ea \right)
\ee
which can also be re-written as  ({\it c.f.} CKS-SM (11))
\be
\left( \ba{c} \gamma_{\vec q} \\ \gamma_{-\vec q}^{\dagger} \ea \right) 
= \mathcal{M} \left( \ba{c} \xi_{\vec q} \\  \Pi _{\vec q} \ea \right)
\ee
with thus
\be
\mathcal{T}^{-1} =  \frac{ \mathcal{M} }{\sqrt{2}} \left( \ba{cc} 1 & 1 \\ -i  & i \ea \right) 
\label{TM}
\ee

$\mathcal{T}$ satisfies  (MSM (10))
\be
\mathcal{T} \left( \ba{cc} 1 & 0 \\ 0 & -1 \ea \right) \mathcal{T}^{\dagger} =  \left( \ba{cc} 1 & 0 \\ 0 & -1 \ea \right)
\ee
and hence also the same equation with $\mathcal{T}$ replaced by $\mathcal{T}^{-1}$.  Eq (\ref{TM}) then shows that
\be
i \mathcal{M}  \left( \ba{cc} 0 & 1 \\ -1 & 0  \ea \right)  \mathcal{M}^{\dagger} =  \left( \ba{cc} 1 & 0 \\ 0 & -1 \ea \right) 
\label{CKS7}
\ee
thus equivalently CKS-SM (7).

Since we write the equation of motion for the operators $\xi_{\vec q}, \Pi_{\vec q}$ in the form eq (\ref{matrix2})
and we have defined $(u,v)$ via (\ref{uv}), comparison with CKS-SM (4) and (5) shows that $(u, v)$ are just the rows of the
matrix $\mathcal{M}$.  
The normalization condition
\be
i ( \vec u \cdot \vec v^* - \vec v \cdot \vec u^*) = 1
\label{norm}
\ee
follows from (\ref{CKS7}).

\section{Berry Curvature }\label{curvature}

The Berry curvature for a given band $n$  is given in  MSM's eq. (34):
\be
\Omega_{B, j} = i \epsilon_{jkl} \left[
  \left( \ba{cc} 1 & 0 \\ 0 & -1 \ea \right) \frac{ \partial  \mathcal{T}^{\dagger}}{\partial q_k}   \left( \ba{cc} 1 & 0 \\ 0 & -1 \ea \right)
 \frac{ \partial  \mathcal{T}}{\partial q_l}  \right]_{nn}
\label{MSM34}
\ee
Eq.  (\ref{CKS7}) implies that 
\be
 \left( \ba{cc} 1 & 0 \\ 0 & -1 \ea \right) \mathcal{T}^{\dagger}  \left( \ba{cc} 1 & 0 \\ 0 & -1 \ea \right) 
= \frac{ \mathcal{M} }{\sqrt{2}} \left( \ba{cc} 1 & 1 \\ -i  & i \ea \right)
\ee
Substituting this into eq. (\ref{MSM34}) we get
\be
\Omega_{B, j} =  \epsilon_{jkl} \left[
  \frac{ \partial  \mathcal{M}}{\partial q_k}   \left( \ba{cc} 0 & - 1 \\ 1 &  0 \ea \right)
 \frac{ \partial  \mathcal{M}^{\dagger}}{\partial q_l}  \left( \ba{cc} 1 & 0 \\ 0 & -1 \ea \right) \right]_{nn}
\ee
Using that the rows of $\mathcal{M}$ are $(\vec u, \vec v)$, we obtain the Berry curvature
\be
\Omega_{B, j}   =   - \epsilon_{jkl}  \left( \frac{\partial \vec u}{ \partial q_k}  \cdot  \frac{\partial \vec v^*}{ \partial q_l}
- \frac{\partial \vec v}{ \partial q_k}  \cdot  \frac{\partial \vec u^*}{ \partial q_l} \right) 
\label{CKS13}
\ee
Note that the right hand side of this equation is real \cite{extra-i}. 

The Berry curvature can be easily evaluated using  eq. (\ref{CKS13}).  We display some formulas for the
transverse modes, where $\vec u  = u_\theta \hat \theta + u_\phi \hat \phi $,  $\vec v  = v_\theta \hat \theta + v_\phi \hat \phi$, with 
$u_{\theta}$, .. $v_{\phi}$ depending only on $q$, $\theta$ but not $\phi$:

\begin{widetext}
\be
\vec \Omega_B \cdot \hat q = - \frac{2}{q^2} {\rm Re} \left[
\left( u_\theta v^*_{\phi}- u_{\phi} v_\theta^* \right)  + 
\frac{ \cos \theta }{\sin \theta} \left( - \frac{\partial}{\partial \theta} ( u_\theta v_\phi^*) + \frac{\partial}{\partial \theta} ( u_\phi v_\theta^*) \right) 
 \right] 
\label{Bq}
\ee
\be
\vec \Omega_B \cdot \hat \theta =  -  \frac{2 \cos \theta}{ q \sin \theta} {\rm Re}
\left[ \frac{\partial}{\partial q} (  u_\theta v^*_{\phi} - u_{\phi} v^*_\theta )  \right]
\label{Btheta}
\ee
\be
\vec \Omega_B \cdot \hat \phi =  - \frac{2}{q} {\rm Re} \left[
\frac{\partial u_\theta}{\partial q} \frac{\partial v^*_\theta}{\partial \theta} + \frac{\partial u_\phi}{\partial q} \frac{\partial v^*_\phi}{\partial \theta}
- \frac{\partial u_\theta}{\partial \theta} \frac{\partial v^*_\theta}{\partial q} + \frac{\partial u_\phi}{\partial \theta} \frac{\partial v^*_\phi}{\partial q}
\right]
\label{Bphi} 
\ee

In eqs. (\ref{Bq}-\ref{Bphi}) we have dropped the subscripts $q$ of $\theta_q$ to simplify the notation. 

\end{widetext}



\begin{thebibliography}{99}
\bibitem{eta} For Sr$_2$IrO$_4$,  the separation between the $j_{eff} = 1/2$ and $3/2$ manifold is  several hundred meV according to \cite{Kim08}.
\bibitem{notei}  Our choice of phase factors in eq (\ref{ps}) contains an overall  extra factor $-i$ compared with \cite{Shitade09}.
This is to gaurantee the time-reversal symmetry mentioned.  See also the remark \cite{noteLs} below.
\bibitem{noteLs} The choice in \cite{JK09}
would correspond to removing the factors $\mp i$ in eq (\ref{ps}), we would have instead $L_{z} = - \frac{2}{3} \tau_{z}$ but
$L_{x,y} = + \frac{2}{3} \tau_{x,y}$ and similarly for the spin.  Similary the choice in \cite{Kim09} would not yield a simple vector relation as in eq (\ref{Lstau}).
\bibitem{drop}  The isotropic compression can cause an overall shift in energy but that would be common to both energy levels in eq (\ref{Psi}), while
the anisotropic can cause virtual transition to higher energy levels.  In neither case can these generate a linear coupling to $\vec \tau$ or $\vec B$, which
are vectors.  We shall be confining ourselves to effects that are first order in $\vec B$.
\bibitem{expand} $\vec \Omega_B \cdot \hat \phi$ vanishes by symmetry, but 
Eq (\ref{theta0}) itself is not sufficient to evaluate the other two components since information near small and finite  $\theta$'s
are needed.  At small $\theta$ and writing the correction factors as $ ( 1 + c_1 \theta^2)$ with $c_{1,2,3,4}$ etc. for
the four components in eq (\ref{theta0}) and $ c_1+c_2+c_3+c_4 = 0$ due to normalization. 
This implies the results given in text.
\bibitem{estimate} This intrinsic contribution is a sum over phonon modes weighted by thermal factors
for two circularly polarized modes with opposite Berry curvatures hence opposite contributions to $\kappa_{xy}$. 
Note that, due to the remark after eq (\ref{Z}), the lower frequency mode for both
$q_z { > \atop < } 0$  have $\vec \Omega_B \cdot \hat z < 0$ and so adds.
 Hence
$\kappa_{xy} / T \sim q^3 \times ( 1 / q^2 ) \times ( \hbar \delta \omega /  k_B T) ( k_B^2 / \hbar) $ where
the first factor is from the $q$ integral, the second from the Berry curvature, and the third
from the difference in occupation numbers of the two modes, and each of these factors can be replaced by
the typical value at a given temperature $T$.  This estimate also follows from dimensional analysis. 
($\kappa_{xy}/T$ has dimension of inverse length apart from the factor $k_B^2/\hbar$). 
A more detailed calculation using the formulas in \cite{Qin12,Matsumoto} yields the same result except a numerical factor of order 1.
\bibitem{extra-i}   CKS-SM (13) seems to contain some typos.  Their expression is imaginary.
\end{thebibliography}
\end{document}